\documentclass{aa}
\usepackage{graphicx}
\usepackage{natbib}
\usepackage{txfonts}

\begin{document}

\bibliographystyle{aa}
\def\ctq{CTQ~414}

\def\kmsmpc{km~s$^{-1}$~Mpc$^{-1}$}
\def\rxj{RX~J0911.4+0551} 
\def\he{HE~0230-2130} 
\def\h1413{H~1413+117} 
\def\b1359{B~1359+154} 
\def\ho{H$_0$}
\def\sex{SExtractor}

  \title{Integral field spectroscopy of four lensed quasars: analysis of their neighborhood and evidence for microlensing}

  \subtitle{}
   
  \author{T. Anguita\inst{1} \and C. Faure\inst{1} \and A. Yonehara\inst{1,2} \and J. Wambsganss\inst{1} \and  J.-P. Kneib\inst{3} \and G. Covone\inst{4} \and D. Alloin\inst{5}}

\institute{Astronomisches Rechen-Institut, Zentrum f\"{u}r Astronomie der Universit\"{a}t Heidelberg, M\"{o}nchhofstrasse 12-14, 69120 Heidelberg, Germany.
	\and
JSPS Postdoctoral Fellowships for Research Abroad
         \and
Laboratoire d'Astrophysique de Marseille, Traverse du Siphon, Les Trois Lucs, BP8-13376 Marseille Cedex 12, France
	\and
INAF, Osservatorio Astronomico di Capodimonte, Naples, Italy
	\and
AIM, CEA/DSM-CNRS-Universit\'{e} Paris 7, SAp, B\^{a}t. 709, CE Saclay, l'Orme des Merisiers, 91191 Gif-sur-Yvette Cedex, France}

\authorrunning{T. Anguita et al. }

\titlerunning{Integral field spectroscopy of lensed quasars}

\abstract{Gravitationally lensed quasars constitute an independent tool to derive the Hubble constant through the time-delays of their different images; they offer as well the opportunity to study the mass distribution and interstellar medium of their lensing galaxies and, through microlensing due to stars in their lensing galaxy, they also allow one to study details of the emitting source.}{For such studies, one needs to have an excellent knowledge of the close environment of the lensed images in order to model the lensing potential: this means, ideally, observational data over a large field-of-view and spectroscopy at high spatial resolution.}{We present VIMOS integral field observations of 27\arcsec$\times$27\arcsec\ fields of view around four lensed quasars: HE~0230-2130, RX~J0911.4+0551, H~1413+117 and B~1359+154. Using the low, medium and high resolution modes, we study the quasar images and the quasar environments. Because of the complexity of the reduction of integral field datasets, we provide a detailed report of the data reduction.}{Comparison between the quasar spectra of the different images reveals differences for \he, \rxj\ and \h1413: flux ratios between the images of the same quasar are different when measured in the emission lines and in the continuum. We have also measured the redshifts of galaxies in the neighborhood of \he\ and \rxj\ which possibly contribute to the total lensing potential, given their close proximity to the line-of-sight toward the quasars.}{A careful analysis reveals that microlensing is the most natural explanation for the (de)magnification of the continuum emitting region of the background sources. In HE~0230-2130, image D is likely to be affected by microlensing magnification; in RX~J0911.4+0551, images A1 and A3 are likely to be modified by microlensing de-magnification and in \h1413\, at least image D is affected by microlensing. We have not been able to recover any microlensing information or galaxies close to the line-of-sight in the neighborhood of \b1359.}


\date{Accepted 07/01/2008}

\maketitle

\section{Introduction}

Gravitationally lensed quasars are a very useful tool to study the intergalactic medium, the mass distribution and substructures of the lensing galaxies \cite[]{nadeau91, falco99, wucknitz03, yonehara03}. They are used in the quest for cosmological parameters \cite[]{bradac04,meneghetti05,taylor05,mortsell06}, in particular the Hubble constant ($H_0$) through time-delay measurements \cite[]{refsdal64}. A list of measured time-delays currently available has been established \cite []{oguri07}. The Hubble constant can be derived using the time-delays between the images of the quasar and a model of the mass distribution of the lens, for which the amplification is directly comparable to the observed flux ratio between the images. However, lensing galaxies do not provide ``smooth'' potentials, but rather consist of stars. Therefore, there is often a supplementary amplification caused by stars in the lensing galaxies close to the line-of-sight. This phenomenon is known as microlensing \cite[]{chang79,wambsganss91,schechter02,SKW06}. Its study may allow us to statistically understand the flux ratios in order to use them to constrain the ``macro'' model of the lens.

 Microlensing also provides valuable information about the background source object and the lens itself. By coupling photometric or spectroscopic observations to theoretical microlensing simulations (analytical or numerical), one is able to derive the size \cite[e.g.,][]{wambsganss90,yonehara01,kochanek06} and shape \cite[e.g.,][]{mineshige99} of the background source, as well as the transverse velocity  \cite[e.g.,][]{GM05} and mean mass of the microlenses \cite[e.g.,][]{kochanek04}.

The programme we are reporting upon in this paper was conducted at ESO in period 74 (proposal IDs: 074.A-0152(A) and 074.A-0152(B), PI: C. Faure) and intended to spectroscopically study in detail four gravitationally lensed quasars (\he, \rxj, H~1413+117 and B~1359+154), search for the occurrence of microlensing and study their close neighborhood.

 Using the integral field spectroscopy mode of the Visible Multi Object Spectrograph (VIMOS) instrument installed on the ESO/VLT-UT2 telescope in Cerro Paranal (Chile), we have obtained 3D spectroscopy of the fields around the four targets. These are briefly introduced in Sect. \ref{target}. The dataset is presented in Sect. \ref{observations}. The data reduction of the integral field spectroscopy, in its low, medium and high resolution modes, is detailed in Sect. \ref{observation}. The identification and extraction of the object spectra in the quasar fields, together with their analysis, are reported upon in Sect. \ref{extraction}. Finally, the microlensing analysis is presented and discussed in Sect. \ref{discussion}, followed by a summary and conclusions in Sect. \ref{conclusion}.

\section{The targets}\label{target}

The lensed quasar targets were selected based on the fact that the presence of a galaxy cluster/group was either known or suspected in the line-of-sight of the quasars. Furthermore, for \rxj\ and \h1413, microlensing activity was suspected \cite[]{bade97,angonin90}.

\subsection{The quadruple quasar \he}

The quadruple lensed quasar \he\ was discovered by \cite{wisotzki99} in the course of the Hamburg/ESO survey \cite[]{wisotzki96}. The quasar is at redshift z=2.16. The separation between the four quasar images ranges from  $\Delta_{AB}$=0.74\arcsec\, to  $\Delta_{AD}$=2.1\arcsec. A galaxy overdensity was discovered  40\arcsec\ south-west of the quasar images  \cite[]{faure04}. \cite{eigenbrod06} identified the redshift of the two main lensing galaxies that lie between the quasar images respectively at z$_{LG1}$=0.523 $\pm$ 0.001 and z$_{LG1}$=0.526 $\pm$ 0.002. The images of the lensed quasar are displayed in the left insert of the right panel in Fig. \ref{hefield}.

\subsection{The quadruple quasar \rxj}

The quasar  \rxj\, was originally selected from the ROSAT All-Sky Survey (RASS) \cite[]{bade95}. It was classified by \cite{bade97} as a multiply imaged quasar with at least three images at redshift z=2.80. In an analysis with higher resolution imaging from the Nordic Optical Telescope (NOT, La Palma) observations, \cite{burud98} discovered a fourth image of the quasar.  The lensed quasar is described as three very close images (A1, A2 and A3, with $\Delta_{A1A2}$=0.48'', $\Delta_{A1A3}$=0.96'') and an isolated fourth image (B) located $\Delta_{A1B}$=3.05'' west of image A1. The lensing system comprises an elongated lensing galaxy located at $\Delta_{A1G}$=0.86'' and a galaxy cluster centered at a distance of 38'' from the quasar images. The redshift of the lensing galaxy was measured from Keck observations: z$_l$=0.769 \cite[see][]{KCH00}. \cite{KCH00} characterized the galaxy cluster located south-west of the quasar images at z=0.769$\pm$0.002, based on the spectroscopic measurements of 24 galaxy members. The measured velocity dispersion of the galaxy cluster is of: $\sigma$= $836^{+180}_{-200}$ ~km~s$^{-1}$.
The quasar images and the lensing galaxies are displayed in Fig. \ref{rxfield}.

\subsection{The quadruple quasar \h1413}

The quadruply imaged quasar \h1413\ was first identified as a gravitationally lensed quasar by \cite{magain88}. It is one of the rare multiply imaged Broad Absorption Line (BAL) quasars. The quasar is at a redshift z=2.55, while the spectroscopic redshift of the lensing galaxy is still unknown. The presence of a galaxy cluster in the direction of the quasar was first suspected by \cite{kneib98} and confirmed by \cite{faure04}, at a photometric redshift z$\sim0.8\pm0.3$ (using up to six photometric bands). Additional absorption systems were identified in the BAL quasar spectrum at redshifts z=1.66 and z=1.44 \cite[]{magain88}. The different components of the system are displayed in Fig. \ref{h1413field}.

\subsection{The multiple quasar \b1359}
	
	The lensed quasar \b1359\ was discovered as part of the CLASS survey and discussed by \cite{myers99}. The quasar is at redshift z=3.24 and has a maximum image separation of 1.7''. As identified by \cite{rusin01}, this system shows at least six images of a radio source and its star forming host galaxy. The main share of the lensing potential is produced by three lensing galaxies inside the Einstein ring with currently no spectroscopic redshift available; however, the three galaxies have been identified as the core of a galaxy group possibly at redshift z$\simeq$1 based on optical colors and mass estimates derived from models of the system \cite[]{rusin01}.

\section{The VIMOS dataset}\label{observations}
This new set of observations was collected in service mode between October 2004 and March 2005. The data were taken with the Integral Field Unit (IFU) using the low, medium and high spectral resolution modes.

\subsection{The VIMOS integral field unit}

The  VIMOS integral field unit (IFU)  consists  of  6400 fibers coupled to microlenses, divided into four quadrants. Each quadrant is made of four sets of 400 fibers, called ``pseudo-slits''. Each pseudo-slit corresponds to a 20$\times$20~pixels area  within each quadrant of the IFU head (80$\times$80 pixels in total).

The instrument can work in three resolution modes: low resolution (LR, spectral resolution=210-260, dispersion=5.3~\AA{}~pix$^{-1}$),  medium resolution (MR, spectral resolution=580-720, dispersion=2.5~\AA{}~pix$^{-1}$) and high resolution (HR, spectral resolution=2500-3100, dispersion=0.6~\AA{}~pix$^{-1}$). In the LR mode all fibers are used whereas in the MR and HR modes only one pseudo-slit is used per quadrant. Therefore, the size of the field-of-view depends on the spatial sampling. We have selected the sampling so as to cover a 27'' $\times$ 27'' field size (0.33''/fiber in LR and 0.67''/fiber in MR/HR). The targets were observed through the LR-blue, MR and/or HR-red/orange grisms\footnote{For HR, the observations through quadrants 1, 2 and 3 were obtained with the red filter, while there was no red grism for quadrant 4. Here we used the orange grism instead, see VIMOS Handbook 30 June 2005.}.

We have coupled the  LR-blue grism with the OS-blue filter to cover a wavelength range of  $\lambda$=3700-6700 \AA{}. The MR- and HR-red grism observations were coupled with  the GG475 filter to cover a wavelength range of $\lambda$=5000-10000 \AA{} and $\lambda$=6350-8600 \AA{} respectively, while the HR-orange grism  was coupled  with the GG435 filter to cover a wavelength range of $\lambda$=5250-7550 \AA{}.

\subsection{The observing runs}

A summary of the observing runs used for this analysis is displayed in Table 1. For \rxj\ and \he\ we have retained only the exposures conducted under a seeing below 1.1'' and an airmass less than 1.7. For \h1413\ and \b1359\ we have considered the observations including at least three quadrants and obtained with an airmass below 1.7, as well. This represents total exposure times of 7 ks in LR and 11 ks in HR for \he, of 7 ks in LR and 9 ks in HR for \rxj, of 7 ks in LR and 9 ks in MR for \h1413\ and of 11 ks in MR for \b1359. 

\begin{table}
\renewcommand{\arraystretch}{1.0}
\centering
\begin{center}
\caption{Observation log (all observations performed between October 2004 and March 2005).}

{\scriptsize
\begin{tabular}{l l l l l l}
\hline
\hline
Object&  Mode   & Night &FWHM&Air &Exp\\
      &         &       &  ''  & Mass        &sec\\
\hline
\hline      
\he	&LR Blue&	Nov. 17&	0.60&	1.24&	2355\\
	&	&Nov. 17&	0.58&	1.50&	2355\\
	&	&Dec. 5&	1.08&	1.02&	2355\\
\he	&HR Red	&Oct. 16&	0.60&	1.18&	2205\\
	&	&Oct. 16&	0.72&	1.07&	2205\\
	&	&Oct. 16&	0.87&	1.02&	2205\\
	&	&Oct. 16&	0.61&	1.09&	2205\\
	&	&Nov. 13&	0.54&	1.24&	2205\\
\rxj	&LR Blue&	Dec. 8&	0.58&	1.28&	2355\\
	&	&Dec. 9&	0.67&	1.31&	2355\\
	&	&Dec. 15&	0.83&	1.29&	2355\\
\rxj	&HR Red	&Dec. 5&	0.47&	1.25&	2205\\
	&	&Dec. 5&	0.46&	1.41&	2205\\
	&	&Dec. 10&	0.52&	1.24&	2205\\
	&	&Dec. 10&	0.50&	1.40&	2205\\
\h1413	&LR Blue&	Mar. 18&	1.27&	1.25&	2355\\
	&	&Mar. 18&	1.13&	1.26&	2355\\
	&	&Mar. 18&	1.21&	1.64&	2355\\
\h1413	&MR Orange	&Mar .17&	0.83&	1.24&	2205\\
	&	&Mar. 17&	1.22&	1.29&	2205\\
	&	&Mar. 17&	1.03&	1.42&	2205\\
	&	&Mar.17&	1.12&	1.68&	2205\\
\b1359	&MR Orange	&Feb .13&	1.40&	1.24&	2205\\
	&	&Mar. 4&	0.81&	1.33&	2205\\
	&	&Mar. 15&	0.59&	1.30&	2205\\
	&	&Mar.15&	0.39&	1.33&	2205\\
        &	&Mar.16&	0.88&	1.38&	2205\\
\hline
\end{tabular}
}
\end{center}
\end{table}

\section{The data reduction}\label{observation}

The data reduction has been performed mainly using the VIMOS Interactive Pipeline \& Graphical Interface \cite[VIPGI][]{scodeggio05,zanichelli05}. While most information concerning the data processing can be retrieved from the VIPGI Cookbook \cite[]{garilli06} and some important issues are discussed in \cite{covone06}, we find  it necessary to provide additional details on some of the most complex reduction steps in order to make the VIMOS IFU data reduction more transparent for future VIPGI users.

\subsection{First adjustments}
	The location of the object spectra and the wavelength calibration both require an accurate description of the instrumental distortions. The pattern of these distortions is provided in the data header. However, the instrument flexures change with the pointing direction and therefore these values can only be used as ``first guesses'' for a further improvement by polynomial fits. The VIPGI software offers the opportunity to tailor these first guesses separately for each frame in the observation run.

 During this step, individual fibers can be flagged as ``dead" if necessary (i.e. zero fiber transmission/efficiency). The flagged fibers will not be considered in any of the subsequent fittings. If the flagging is not made properly, it yields a distorted image of the astronomical object as illustrated for example in Fig. \ref{wrong}. A quality control routine allows us to check the adjustments in the different science exposures. The corrected first guesses are then applied to the entire scientific and calibration dataset.

\addtocounter{figure}{3}
\begin{figure}
 \centering
 \includegraphics[width=6cm,bb=0 0 626 298]{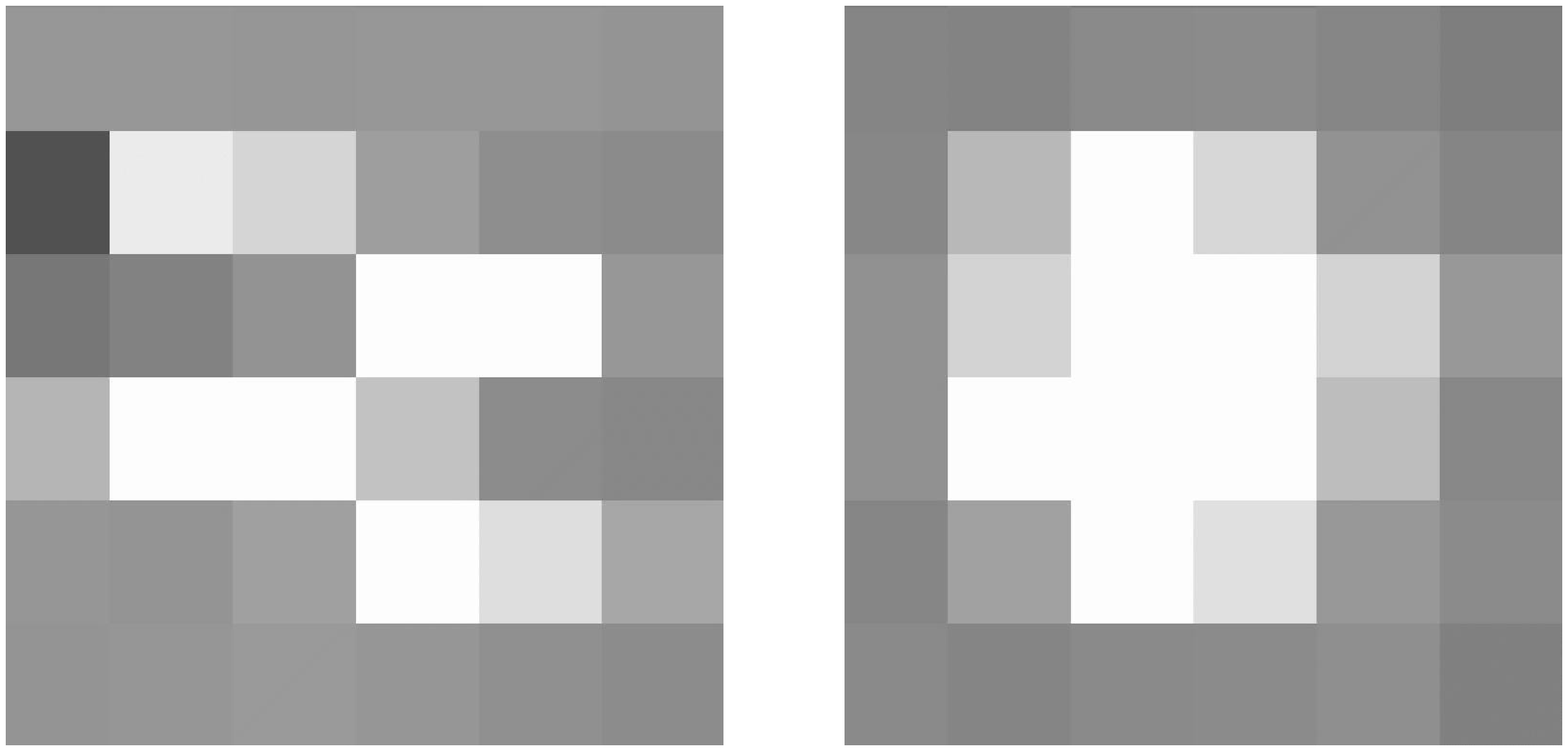}
\caption{\label{wrong} Resulting images of a point-like object in the HR mode for incorrect (left panel) and correct (right panel) flagging of the fibers. In the left panel dead fibers have not been flagged, hence during the first guess adjustment, the software assigns to these fibers the flux of the adjacent fiber. }
\label{fig:1}
\end{figure}

In the second quadrant of the LR mode, the first 80 fibers were flagged as dead because of their low transmission in all frames.

\subsection{Spectra tracing and extraction}
After the first guess correction, one needs to trace the spectra in the different exposures. To do so, we use a high signal flat field exposure obtained immediately after the science exposure. In the LR mode, we use the global fit parameters proposed in the image headers, as they allow a good tracing; for the pixel binning we use 20 pixels to trace the spectra from their start to their end. For the MR/HR frames, better results are obtained ignoring the global fit distortion model: the IFU binning then needs to be larger than 20 pixels (the best fit was obtained for $\sim$50 pixels). From this fit, a table of spectra positions is built. The position table has to be visually inspected to check whether the spectra are correctly traced.

The inverse dispersion solution is calculated next in the ``create master lamp'' routine. This is done by fitting a third order polynomial to the lamp frame, starting from the first guesses. The inverse dispersion solution was checked visually over the raw frames as well as through the distribution of the root mean square (rms) residuals of the fit. The rms residuals are of about 0.7 \AA{} $\pm$ 0.7 \AA{} for the LR frames and 0.09 \AA{} $\pm$ 0.06 \AA{} for the MR/HR frames.

Independently, the science frames are corrected by a master-bias and over-scan trimming. Then, the object spectra are extracted from the science exposures. At this stage, cosmic rays are detected and eliminated and the wavelength calibration is applied.

\subsection{Last calibration and exposure combination} \label{lastcalib}

	The final phase of the reduction process includes many steps. First, the software applies a fiber-to-fiber correction in which the transmission of each fiber is calibrated. This fiber-to-fiber correction is calculated using a bright and isolated sky-line present in all fibers. The transmission is corrected so that each fiber has the same flux all along the sky-line. To perform this correction we have selected the 5577 \AA{} sky-line in LR-blue, MR and HR-orange and the 6300 \AA{} skyline in HR-red.

The sky contribution is dealt with in a statistical way \cite[]{scodeggio05} by subtracting the mode spectrum in each pseudo-slit. This gives satisfactory results if at least 50\% of the fibers in the frames relate to pure sky, which is the case for the current dataset.
 The spectro-photometric calibration of the science frames was made using the observation of standard stars, which accurately corrects for the relative flux.

  The information available in the science frame headers concerning the position of the pointing for our data is not accurate enough to be used as a guide for combining the dithered exposures. Therefore, we have computed the shifts between the different exposures using the brightest objects in the frame. This is a delicate step because of a Point Spread Function (PSF) shape problem which is discussed in Sect. 4.4.

\addtocounter{figure}{-4}

\begin{figure*}[!ht]
	\centering
	\includegraphics[width=12cm,bb=0 0 567 284]{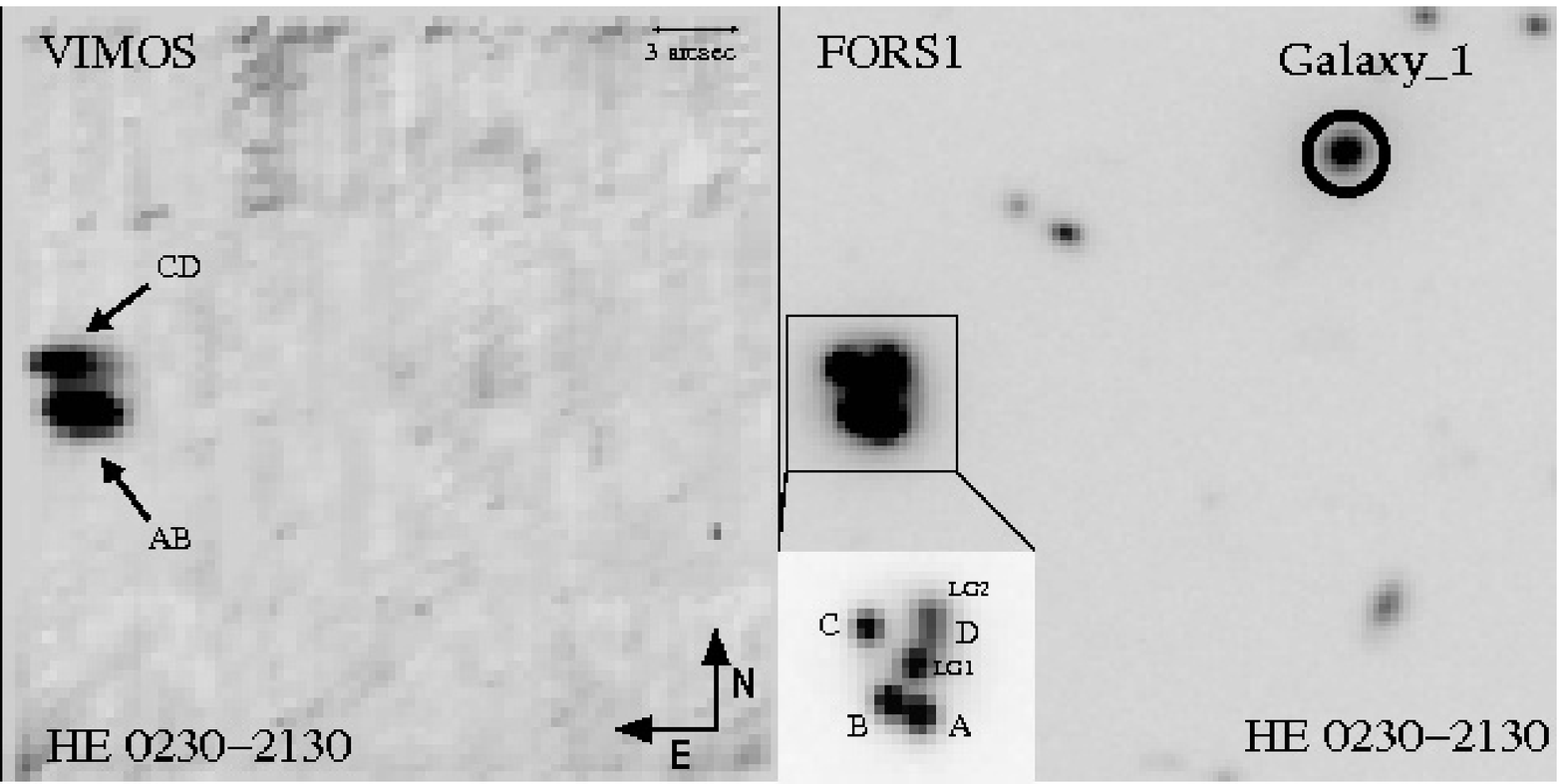}
	\caption{The 27''$\times$27'' field-of-view around the multiple quasar \he: VIMOS IFU LR data cube (left panel) and FORS1 R band image (right panel). The components of the system are labeled as in the text. Orientation and scale are identical in the left and right panels. A FORS1 R band zoom of the system is displayed in the bottom left hand corner of the right panel.}
	\label{hefield}
\end{figure*}

\begin{figure*}[!ht]
	\centering
	 \includegraphics[width=12cm,angle=0,bb=0 0 612 306]{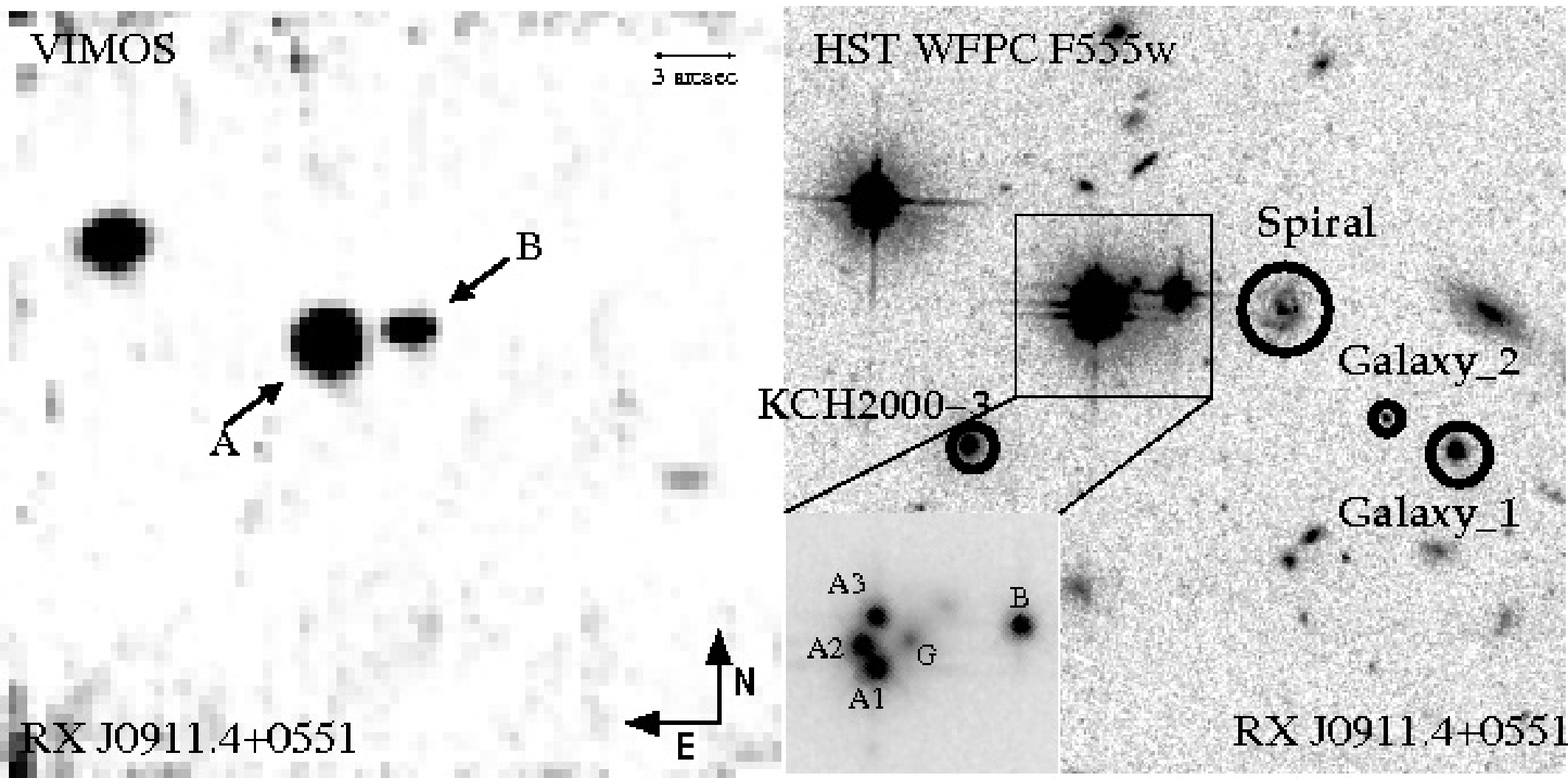}
	\caption{\label{rxfield} The 27''$\times$27'' field-of-view around the multiple quasar \rxj: VIMOS IFU LR data (left panel) and WFPC F555w band image (right panel). The components of the system are labeled as in the text. Orientation and scale are identical in the left and right panels. A WFPC F814w band zoom of the system is displayed in the bottom left hand corner of the right panel.}

\end{figure*}

\begin{figure*}[!ht]
	\centering
	\includegraphics[width=12cm,bb=0 0 567 284]{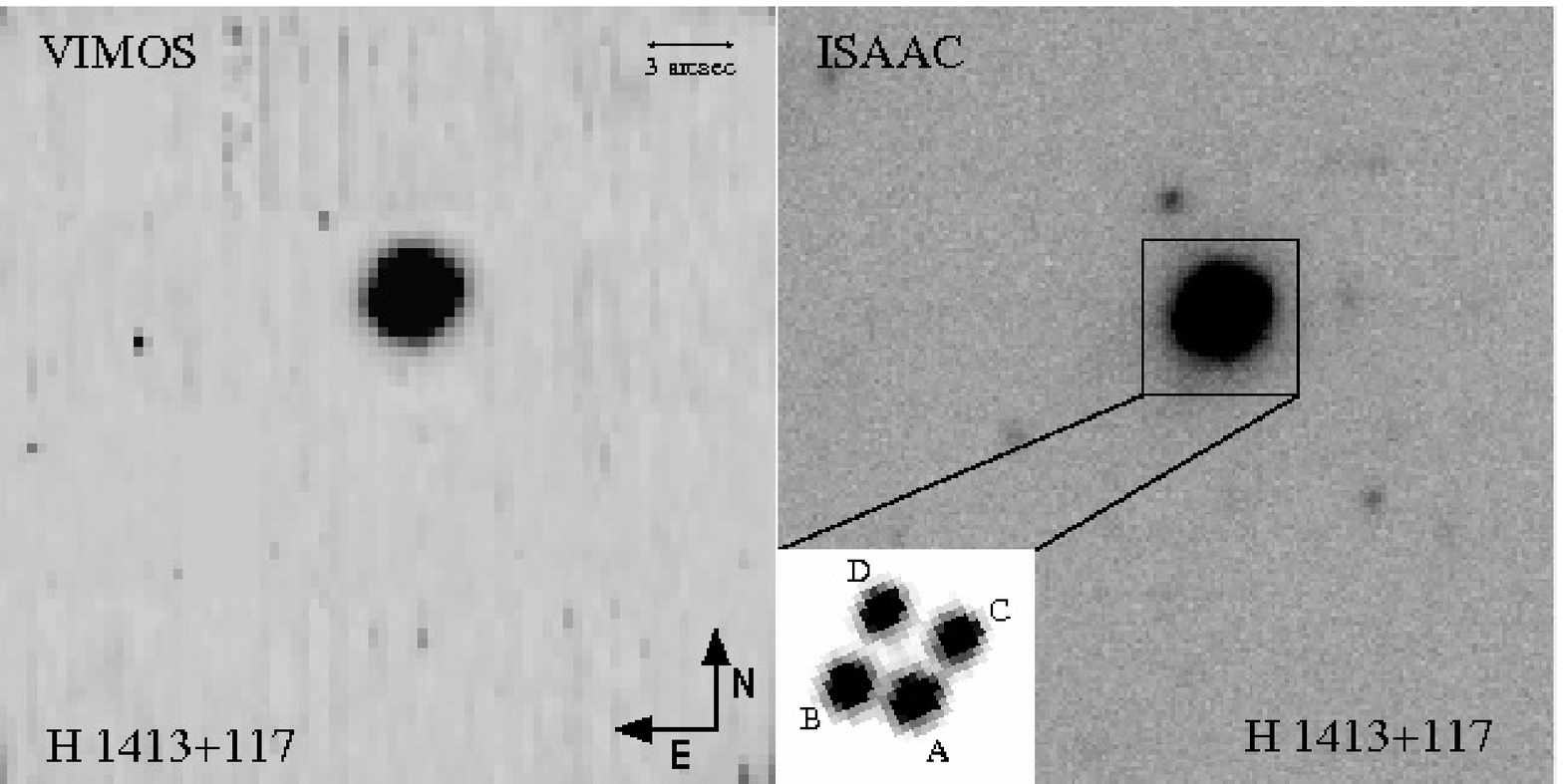}
	\caption{\label{h1413field} The 27''$\times$27'' field-of-view around the multiple quasar \h1413: VIMOS IFU LR data (left panel) and ISAAC J band image (right panel). The components of the system are labeled as in the text. Orientation and scale are identical in the left and right panels. A WFPC F814w band zoom of the system is displayed in the bottom left hand corner of the right panel.}
\end{figure*}

\addtocounter{figure}{1}
Then, the different exposures are combined using a median filter to eliminate the dead pixels and remaining cosmic rays.  The resulting product is a datacube: for each pixel within the 27\arcsec$\times$27\arcsec\ region of the sky, a spectrum is available. During this combination, a fringing correction may be applied.

	The  datacube  is then  corrected for differential atmospheric refraction. Applying a theoretical differential atmospheric refraction correction was attempted but does not give satisfactory results. Instead, we have computed and applied  an empirical differential atmospheric correction: in most cases, the quasars are very bright in the datacube, so we have been able to fit a polynomial to the highest flux position of the quasar image for each wavelength. We then knew the position correction to be applied to each wavelength cut of the datacube.

Images of the fields around \he, \rxj\ and \h1413\ are shown in Figs \ref{hefield}, \ref{rxfield} and \ref{h1413field}. The images were created by adding all the monochromatic slices in the datacube, excluding those corresponding to strong sky emission lines, and excluding the edges of the grisms. In the case of \b1359\ we only have MR observations with one missing quadrant and low signal-to-noise ratio (S/N $\sim$4), hence we could not create such an image.

\subsection{An additional PSF problem with VIMOS}\label{prob}

After the reduction steps and the corresponding checks described above, the PSF in the science target and standard star datacubes still appeared elongated along the East-West direction for all individual exposures. We could not find any VIMOS documentation reporting on this problem; yet, it has been confirmed by ESO staff (private communication) that as the IFU head is placed at the edge of the VIMOS field-of-view, the image of a point source is subject to an aberration and appears slightly elongated in one direction (see for example image B of \rxj\ in the left panel of Fig. 2). This seems to be an identified problem to be mentioned in the VIMOS documentation. Besides the time investment in understanding the reason for the elongation, it is very difficult to quantify how much this issue impairs the image resolution in the datacubes, as this is a combination of other effects as well (seeing, signal-to-noise and dithering).

However, and as shown below, it is still possible to extract relevant spatial information from the data.

\section{Identification of the quasar and galaxy spectra}\label{extraction}
	
A number of tools were used to visualize and analyze VIMOS datacubes. We used mainly the Euro3D software (S\'anchez 2005, http://www.aip.de/Euro3D) that allows us to inspect the datacube, both in an interactive and an automated manner (guidance on the use of the software can be found in the Euro3D documentation).

 We have used existing deep images of the target field-of-views to identify and locate objects that are faint in the VIMOS rebuilt images (27''x27''). We have used VLT/FORS1 R band data to locate objects in the field around \he, HST/WFPC-F555w and I814w for \rxj\ and HST/WFPC-I814w, ISAAC J and ISAAC K for \h1413\ (HST programmes PIs, respectively: J. Hjorth and A. Westphal and VLT programme PI: C. Faure).

The objects for which we have detected a signal are shown and labeled in Figs. \ref{hefield}, \ref{rxfield} and \ref{h1413field} for \he, \rxj\ and \h1413\ respectively. For redshift analysis we use the \textsc{ez}\footnote{Redshift analysis software by the authors of VIPGI: http://cosmos.iasf-milano.inaf.it/pandora/EZ.html} software coupled with spectral templates \cite[]{kinney96}, plus \textsc{iraf} routines. We provide below a detailed analysis of the different spectra.

\subsection{The field of \he }

\subsubsection{The quasar spectra}

\begin{figure}
	\centering
	 \includegraphics[width=8cm,bb=0 0 425 708]{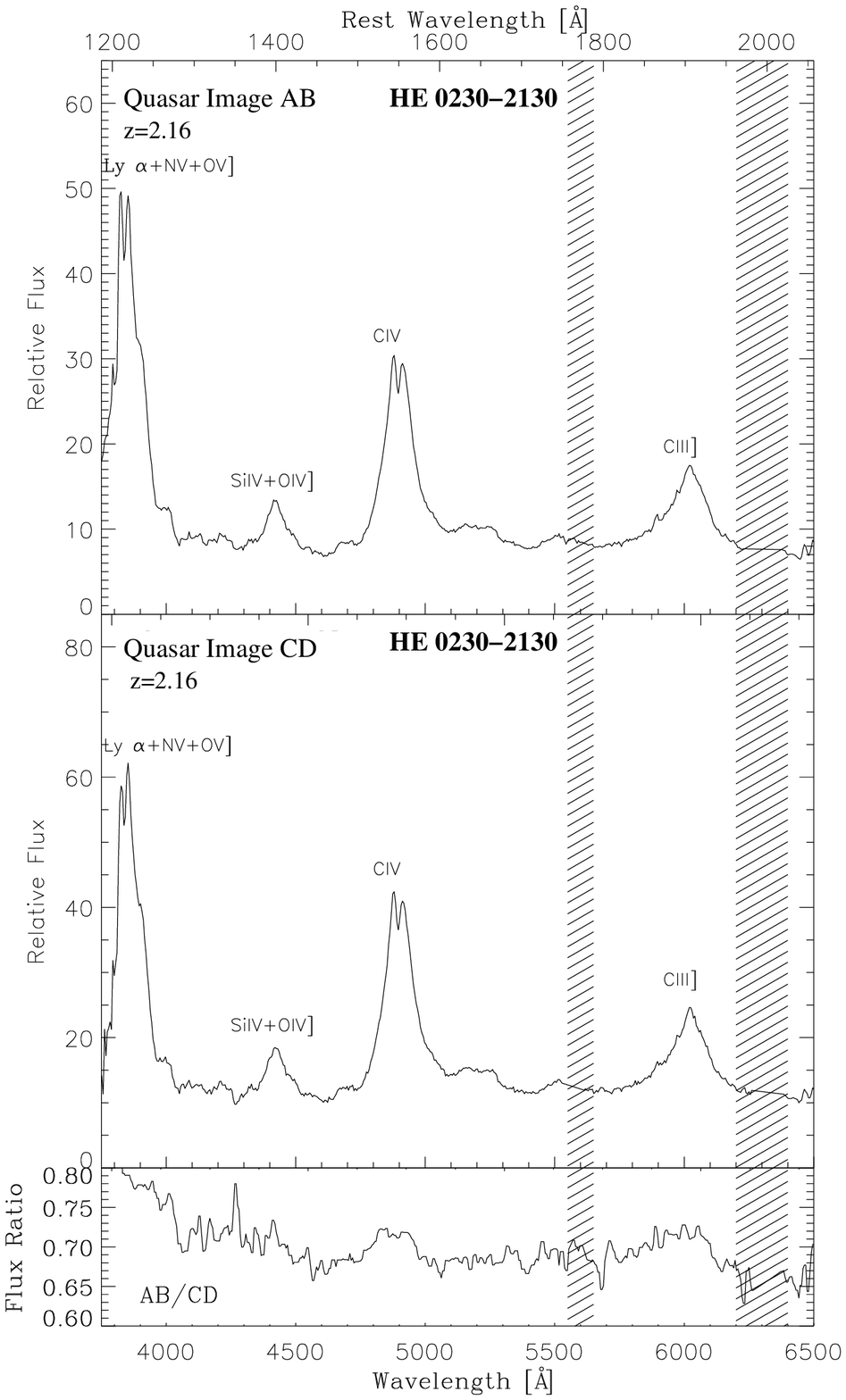}
	\caption{\label{hespeclr}Low resolution spectra for \he. Top panel: combined spectra for images AB. Middle Panel: combined spectra for images CD. Bottom panel: flux ratio between the two spectra. Shaded regions indicate the spectral ranges dominated by intense sky line residuals.}
\end{figure}

	The spectra of the quasar images have a very high signal-to-noise ratio (S/N $\sim$40 in the continuum). However, because of PSF shape problems and of the small angular separation between the quasar images, it is not possible to separate all the spectra of the different quasar images. We have partitioned the images in two groups, A+B (from now on AB) and C+D (from now on CD), which are visually resolved in the final datacube (see Fig. \ref{hefield} for labels). Based on the Lyman $\alpha$(1215\AA{}) + NV(1240\AA{}), SiIV+OIV](1398\AA{}), CIV(1550\AA{}) and CIII](1908\AA{}) emission lines, the spectra from both AB and CD provide a source redshift z=2.163 $\pm$ 0.003, in agreement with \cite{wisotzki99}. In the two spectra an absorption system is detected at the same redshift as the quasar, likely originating from the quasar host galaxy. No other obvious absorption system is detected in the spectra (see Fig. \ref{hespeclr}).

\subsubsection{Galaxies in the field around \he}

	We have extracted spectra at the location of the main lensing galaxies (LG1 and LG2 as shown in Fig. \ref{hefield}) that are highly contaminated by quasar emission. Therefore, a scaled version of the spectrum of the isolated quasar image C has been subtracted. Doing so, we are able to detect absorption lines from both lensing galaxies. We measure the redshift of LG1 at z=0.521$\pm$0.004 and of LG2 at z=0.524$\pm$0.003. These redshifts are in agreement with those obtained by \cite{eigenbrod06} ($z_{\rm LG1}$=0.523$\pm$0.001; $z_{\rm LG2}$=0.526$\pm$0.002).

\begin{figure}
	\centering
	\includegraphics[width=8cm,bb=0 0 504 360]{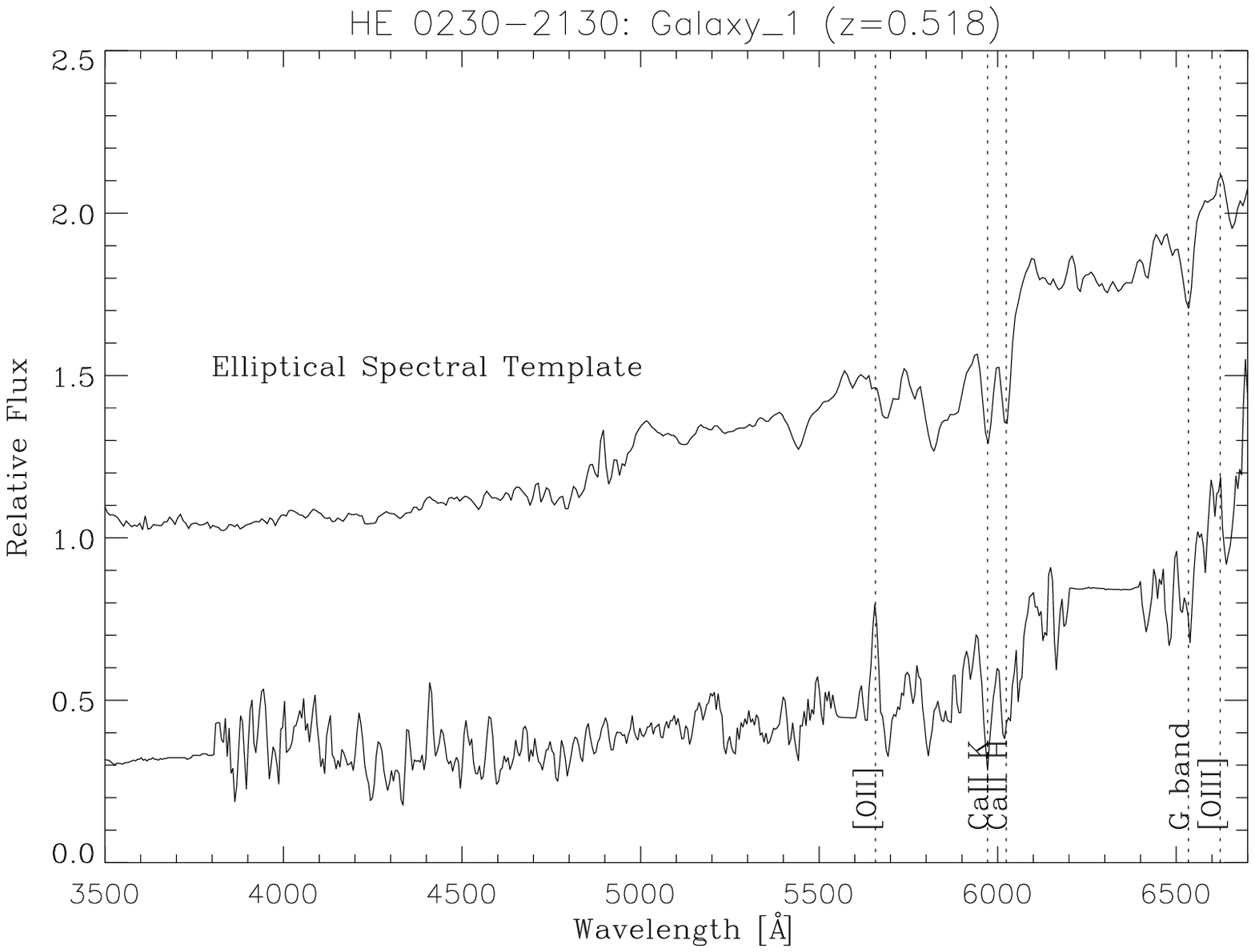}
	\caption{\label{elliphespec}Neighborhood of \he: Low resolution spectrum of the galaxy in the field around \he\ (labeled Galaxy\_1 in Fig. 2) and a scaled version of a template elliptical galaxy spectrum. From the lines shown we measure a redshift of z=$0.518 \pm 0.002$.}
\end{figure}

Another galaxy is seen in the field, north west of the quasar image A ($\Delta \alpha$=-15'', $\Delta \delta$=9.2'', Galaxy\_1 in Fig. \ref{hefield}). Its spectrum shows the CaII K, Ca II H and G-band absorptions, and also the [OIII] and [OII] emission lines (see Fig. \ref{elliphespec}). From these lines we measure a redshift of z=$0.518 \pm 0.002$, thus, this galaxy is probably a member of the group or cluster to which the lensing galaxies LG1 and LG2 belong.

\subsection{The field of \rxj }

\subsubsection{The quasar spectra}
Because of the small separations between images A1, A2 and A3 ($\Delta_{\rm A1A2}=$0.48'', $\Delta_{\rm A1A3}=$0.96'', $\Delta_{\rm A2A3}=$0.61'') and the distorted shape of the PSF, it was not possible to separate the spectra of the three quasar images. Therefore they are treated as a single object called A (see Fig. \ref{rxfield}, left panel).

\begin{figure}
	\centering
	 \includegraphics[width=8cm,bb=0 0 425 708]{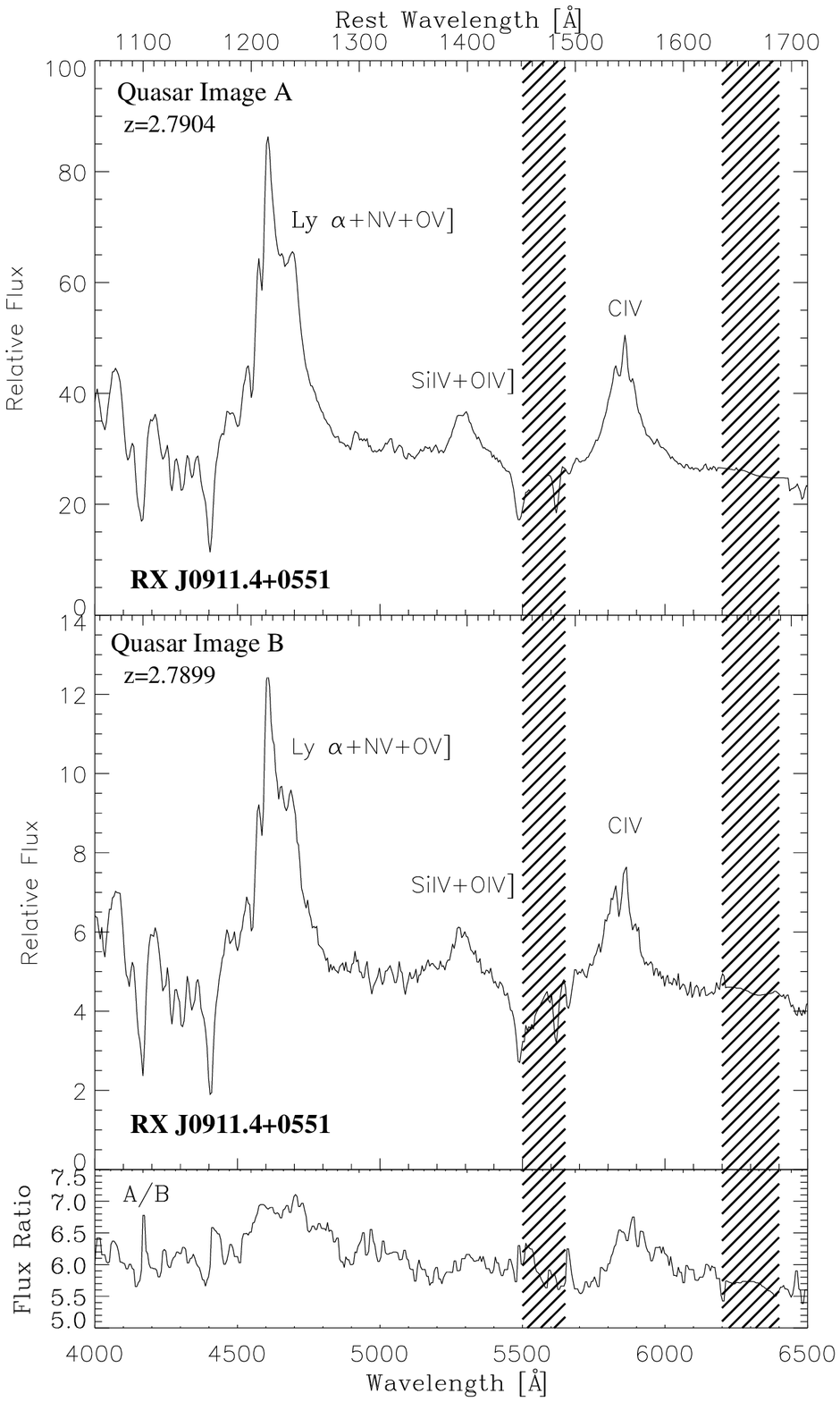}
 	\caption{\label{ratioLRrxj}Low resolution spectra for \rxj. Top panel: summed spectrum for image A. Middle panel: spectrum for image B. Bottom panel: flux ratio between the spectra A and B. Shaded regions indicate spectral ranges dominated by sky residuals.}
\end{figure}

\begin{figure}
	\centering
	 \includegraphics[width=8cm,bb=0 0 425 708]{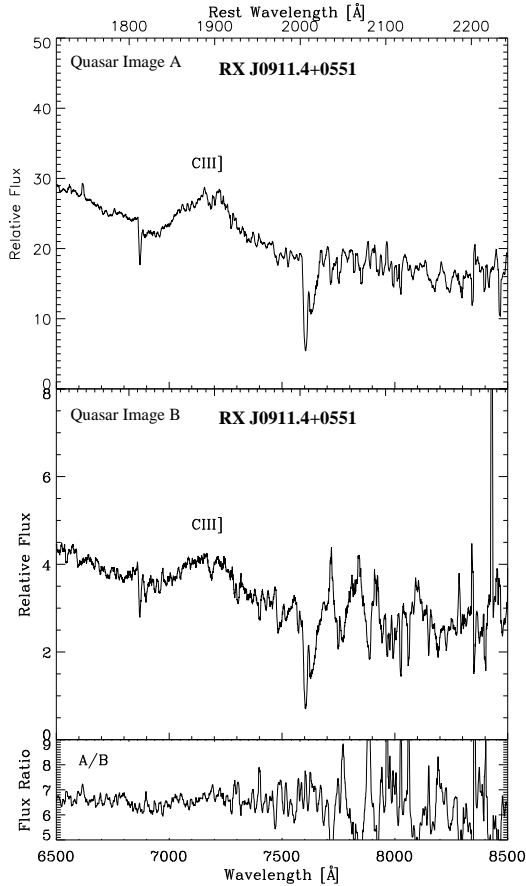}
	\caption{\label{ratioHRrxj}High resolution red spectra for \rxj . Top panel: total spectrum for image A. Middle panel: spectrum for image B. Bottom panel: flux ratio between the two spectra. Sky line residuals dominate the region redder than 7500\AA{}.}
\end{figure}

\begin{table}
\renewcommand{\arraystretch}{1.0}
\centering
\begin{center}
\caption{\label{abslines} Values obtained for the five parameter Gaussian fits of the absorption lines detected in the spectra of \rxj. An asterisk (*) denotes the lines close to the 5577\AA{} sky line.}

{\scriptsize
\begin{tabular}{l l l l l}
\hline
\hline
Redshift&  Line   & Center [\AA{}]& FWHM [\AA{}]& $\sigma$ km${\rm s}^{-1}$\\
\hline
\hline      
z=2.80 &    $Ly\alpha$  & - &    -  &    - \\
  &       $C_{IV}$  & 5840.5  &  15  & 330    \\
  &       $C_{III}$ & 7185.5  &  19  &  340    \\
\hline
z=2.63 &    $Ly\alpha$  & 4404 &30 &   940  \\
  &       $C_{IV}*$  & 5618.5 &    19 &   430  \\
  &       $C_{III}$ & - &    -  &     - \\
\hline
z=2.54 &    $Ly\alpha$ & - &    -  & - \\
  &       $C_{IV}*$  & 5487.5 &   28  &  640  \\
  &       $C_{III}$ & -  &   -  &  - \\
\hline
z=2.45 &    $Ly\alpha$   & 4163 &   29  &  900   \\
  &       $C_{IV}$  & - &  -  &    -\\
  &       $C_{III}$ & - &  -  &   - \\

\hline
\end{tabular}
}
\end{center}
\end{table}

In the LR-blue mode, the A and B quasar spectra exhibit Lyman $\alpha$+NV+OV], SiIV+OIV] and CIV emission lines at redshift z=2.790 $\pm$ 0.009. As already mentioned by \cite{bade97}, we can also detect several absorption lines in both the A and B spectra; CIV and Lyman $\alpha$ at z=$2.42$ (or $\sim-29,000~{\rm km~s^{-1}}$ with respect to the quasar), CIV, SiIV, and Lyman $\alpha$ at z=$2.63$ (or $\sim-13,000~{\rm km~s^{-1}}$ with respect to the quasar), and a few more absorption lines at wavelengths below 4500\AA. An absorption line at 5488\AA{}, if identified as CIV, would imply a redshift of z=2.54 (or $\sim-22,000~{\rm km~s^{-1}}$ with respect to the quasar). It might correspond to the line identified by \cite{bade97} at 5546\AA{} (z=2.57), although its proximity to the 5577\AA{} intense sky line makes its redshift determination uncertain. The absorption lines have a velocity dispersions of $\sigma \le 1000~{\rm km~s^{-1}}$ (see Table \ref{abslines}), hence, the quasar was classified as a mini-BAL quasar \cite[e.g.,][]{chartas01}. The physical origin of the absorption lines is still under discussion; the most likely interpretation is that these lines are due to high-velocity outflows related to the quasar central engine \cite[e.g.,][]{reichard03}.

In the HR-red spectra of the A and B quasar images, we detect the CIII] emission line at the quasar redshift, not seen in the previous spectra of \cite{bade97}. Thanks to the VIMOS high quality data, we also detect a few absorption features on top of the CIV and CIII] emission lines. Their FWHM are smaller than those of the other absorption lines previously indicated (see Table \ref{abslines}). The narrow features could either be produced by outflowing material from the quasar central engine, or signatures of the quasar host galaxy.

In Figs. \ref{ratioLRrxj} and \ref{ratioHRrxj} the spectra of images A and B are shown in low and high resolution modes respectively, as well as the flux ratios between the spectra. Residuals in the flux ratios are discussed in Sect. 6.

\subsubsection{Galaxies in the field around \rxj}

\begin{figure}
	\centering
	\includegraphics[width=8cm,bb=0 0 504 360]{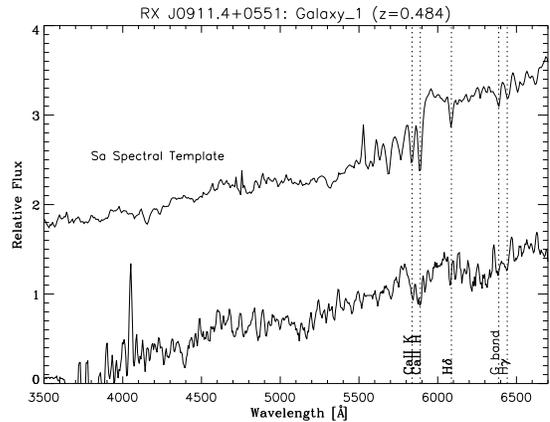}
	\caption{Neighborhood of \rxj: Spectrum of Galaxy\_1 and of an Sa spectral template for comparison. Identified lines are highlighted by dotted straight lines. They yield a redshift z=0.484 $\pm$ 0.001}
	\label{elliprx}
\end{figure}

We are unable to detect any absorption features from the lensing galaxy because of the intense contribution of the three quasar images.

We identify three objects in the field: Galaxy\_1, Galaxy\_2 and Spiral as labeled in Fig. \ref{rxfield}. We are only able to measure the redshift of Galaxy\_1 for which we obtain a value z=0.484 $\pm$ 0.001 (see Fig. \ref{elliprx}), indicating that it is not a member of the cluster found by \cite{KCH00} at z=0.769.

In the HR-red mode, we identify the galaxy labeled KCH2000-3 in Fig. \ref{rxfield}, which had been previously detected by \cite{KCH00} and measured at z=0.7618. Our determination of the redshift gives z=0.762, in good agreement. This galaxy is part of the galaxy group discovered south-west of the quasar.

\subsection{The field around \h1413}

	The quasar images cannot be visually resolved in the two dimensional cuts, either in the LR or in the MR spectroscopy. The low and medium resolution summed spectra for all four quasar images show the Lyman $\alpha$, NV, SiIV emission lines coupled with their broad absorption counterparts and the CIII] line. From the emission lines we measure the quasar redshift at z=2.554 $\pm$ 0.002. 

\begin{figure}
	\centering
	\includegraphics[width=8cm,bb=0 0 504 726]{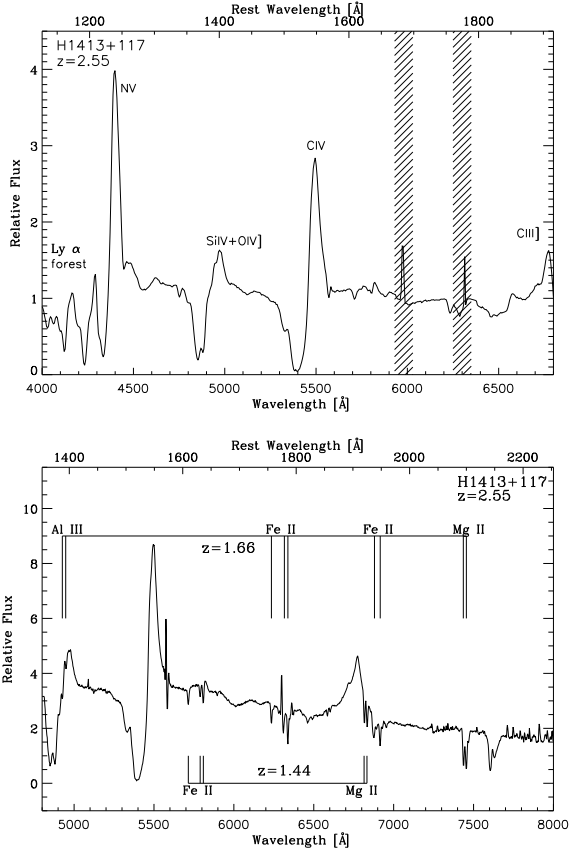}
	\caption{Top panel: LR-blue spectrum for \h1413. The spectrum was reconstructed by adding the contributions from the four quasar images. The dashed areas highlight regions affected by sky subtraction residuals. Bottom panel: MR-orange spectrum for \h1413 and the summed contributions from the four images. The lines identified with the z=1.44 and z=1.66 absorption systems are indicated. Two other doublet systems are seen at $\sim$6450\AA{} and $\sim$6600\AA{}}
	\label{spec1413}
\end{figure}

	The higher spectral resolution and longer wavelength range available in the medium resolution mode allows us to detect the absorption systems already described by \cite{magain88}. Taking advantage of the improved resolution and small width of the absorption lines we have been able to obtain very tight values for their redshifts. In the first absorption system, the FeII triplet ($\sim$5750\AA{}) and the MgII doublet ($\sim$6850\AA{}) yield a redshift z=1.4381$\pm$0.0002; in the second system, the AlIII doublet ($\sim$4950\AA{}), FeII triplet ($\sim$6300\AA{}), the FeII doublet ($\sim$6900\AA{}) and the MgII doublet ($\sim$7450\AA{}) lead to a redshift z=1.657$\pm$0.001. Two other doublets are detected at $\sim$6450\AA{} and $\sim$6600\AA{}; they do not seem to be related and it is hazardous to assign them a specific identification. Low and mid-resolution spectra are displayed in Fig. \ref{spec1413}.

\subsection{The field of \b1359}

	For the quasar system \b1359\, only the MR-mode was used. It being a faint and high redshift quasar, the signal-to-noise ratio is lower than in the other datasets (S/N $\sim4$ in the continuum).

\begin{figure}
	\centering
	\includegraphics[width=8cm,bb=0 0 505 361]{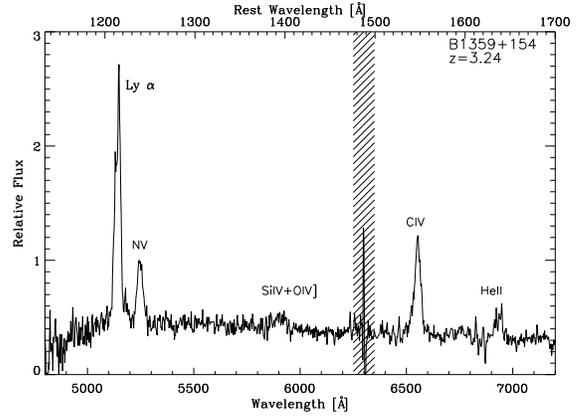}
	\caption{MR-orange spectrum of \b1359 (all spaxels with signal have been summed). Shaded areas correspond to regions affected by sky subtraction residuals.}
	\label{specb1359}
\end{figure}

	However, we can recover a spectrum of the quasar by summing over the spatial elements in the two dimensional wavelength cuts (spaxels) of the different quasar images. The spectrum (displayed in Fig. \ref{specb1359}) exhibits the Lyman $\alpha$, NV, SiIV+OIV], CIV and HeII lines, yielding a redshift z=3.235 $\pm$ 0.003.

	No absorption system is seen in the spectrum. In addition no galaxy is detected in the field. As no spectroscopic signature could be recovered from individual images, no further analysis was possible for this quasar.

\section{Interpretation of the spectral differences between the quasar images}\label{discussion}

\subsection{Spectral differences in \he}

	In order to identify and analyze any differences between the spectra from images AB and CD, we first simply measure the ratio between the two spectra. The ratio varies with wavelength, in particular it is different if measured over the continuum or over the emission lines. Furthermore, a mild slope in the continuum flux ratio is observed. Notice that the edge of the grism ($<$4000\AA{}) is an untrustable region and has been dropped in the analysis.

	To study this phenomenon further, we developed a specific procedure to extract the spectra and spectral energy distribution (SED) of the four individual quasar images: a surface with four Gaussian profiles was fitted to each 2D monochromatic slice of the LR-blue datacube. Distances between the Gaussian centroids correspond to the distances between the quasar images as provided in the CASTLES web page\footnote{http://cfa-www.harvard.edu/castles/}, from HST data. A further constraint is that the FWHM (in both directions: x and y) should be identical for the four profiles at each wavelength cut. The fit was performed using Levenberg-Markwardt routines in an iterative fashion (MPFIT, Craig Markwardt\footnote{http://cow.physics.wisc.edu/$\sim$craigm/idl/}). In Fig. \ref{ratiohead} we show the extracted spectra for images A and D of \he, while their ratio is displayed in Fig. 13 (also reported in Fig. 14). The spectra of the isolated images A and D display emission line differences as well as a mild continuum slope difference. This is not surprising as image D of the quasar is located almost behind one of the lensing galaxies: LG1 (Fig. 2), therefore it is prone to be affected by individual stars and interstellar matter in LG1. Indeed, the flux ratios between images A and B and between images A and C (not shown here) show little or no variation above the noise, contrary to image D.

	Intrinsic quasar variability is a possible, but unlikely, explanation for this phenomenon. Indeed, images A and D are separated by a time-delay and therefore any intrinsic variation of the quasar flux would be reflected at different epochs in images A and D. However, intrinsic variability of the quasar itself would more likely show selective emission line differences rather than similar magnification in all the emission lines. Indeed, different lines being emitted from different regions in the quasar are not likely to vary at the same time \cite[]{kaspi00}. Hence a more plausible explanation for the emission line differences is microlensing by stars in the lensing galaxy, affecting preferentially image D of the quasar.

	 Microlensing is expected to affect the continuum emission more strongly than the broad line emission, because the continuum emitting region is thought to have a size comparable to those of the Einstein rings of stellar microlenses, whereas the broad line region is much larger \cite[e.g.,][]{schneider90,abajas02}. To demonstrate this: based on the virial theorem, the size of the broad line region $(R_{\rm BLR})$ is estimated to be \cite[]{peterson98}:

\begin{figure}
	\centering
	 \includegraphics[width=8cm,bb=0 0 453 341]{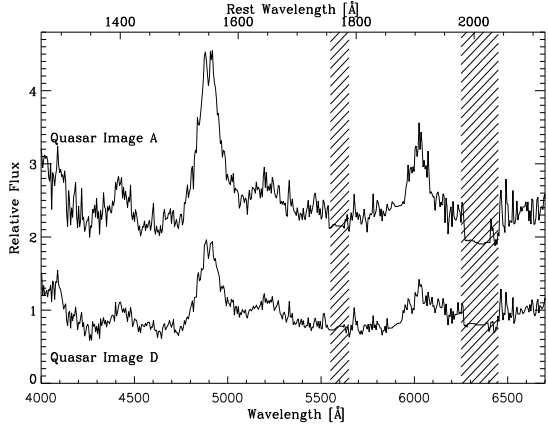}
	\caption{\label{ratiohead}Spectra of the individual images A and D of \he. The normalized spectrum for each image is displayed. The spectrum for image A has been shifted upward by 1.5 flux units for comparison. Shaded areas highlight regions affected by sky subtraction residuals.}
	
\end{figure}

\begin{equation}
R_{\rm BLR}=0.57{\rm pc} \left(\frac{v_{\rm FWHM}}{10^3 {\rm km~s^{-1}}}\right)^{-2} \left(\frac{M_{\rm BH}}{10^8M_{\odot}}\right)
\end{equation}
where $v_{\rm FWHM}$ is the velocity derived from the FWHM of an emission line and $M_{\rm BH}$ is the black hole mass. Assuming a reasonable size for the continuum emitting region $(R_{\rm cont})$, of $1,000$ Schwarzschild radii ($r_s$), we get:

\begin{equation}
 R_{\rm cont}=9.1\times10^{-3} {\rm pc}\left(\frac{M_{\rm BH}}{10^8M_{\odot}}\right)
\end{equation}

It is obvious that $R_{\rm cont} \ll R_{\rm BLR}$, independently of the black hole mass. In this case, the continuum of image D could be magnified by microlensing, making the emission lines ``sink'' in the continuum and, therefore, making them appear small in comparison to the emission lines of image A.

	Could microlensing also explain the slope difference (chromatic effect) seen between the two images A and D? Accretion disk models predict that continuum emission from the central part is bluer than that from the outer part \cite[]{SS73}. In general, the innermost part of an accretion disk (blue-emitting) is more strongly affected by microlensing than the outer part (red-emitting) as the Einstein ring of microlenses matches better the size of the inner accretion disk \cite[]{wambsganss91,yonehara98}: therefore, to first order and in a naive vision, an image affected by microlensing should exhibit a bluer continuum. In image D, the situation is reversed: the quasar image thought to be affected by microlensing (image D) appears redder than the image not affected by microlensing (image A). Let us examine how this could happen.

	In Fig. 13 we illustrate how microlensing can change the slope of the flux ratio between images A and D, by using a simple model for the caustic. We adopt a standard accretion disk model for the continuum emitting region \cite[]{SS73}. The black hole mass is set at $10^8M_{\odot}$, the accretion rate is set fixed at the critical accretion rate and the inner and outer radii are set at  $3r_s$ and $1000r_s$ respectively. Additionally, we assume a single straight line caustic and apply the following approximated formula for magnification ($\mu$) \cite[e.g.,][]{SEF92}: 

\begin{eqnarray}
\mu &=& \left( \frac{x}{x_{\rm scale}} \right)^{-1/2} + \mu_{\rm a}
{\rm ~ ~ ~} (x > 0) \\
&=& \mu_{\rm a} {\rm ~ ~ ~}(x < 0)
\nonumber
\end{eqnarray}

\noindent where $x_{\rm scale}$ is the scale length of the caustic and is chosen to be identical to the Einstein ring radius of a lensing object with mass 1$M_{\odot}$. The parameter $\mu_a$ represents the total magnification except for a pair of micro-images which appear/disappear at the caustic crossing. For image A and image D, the $\mu_a$ values are arbitrarily set to be 3.0 and 2.0 in order to match the observed continuum flux ratio $\sim$1.5 (see Fig. 13).

\begin{figure}
	\centering
	\includegraphics[width=8cm,bb=89 132 511 700]{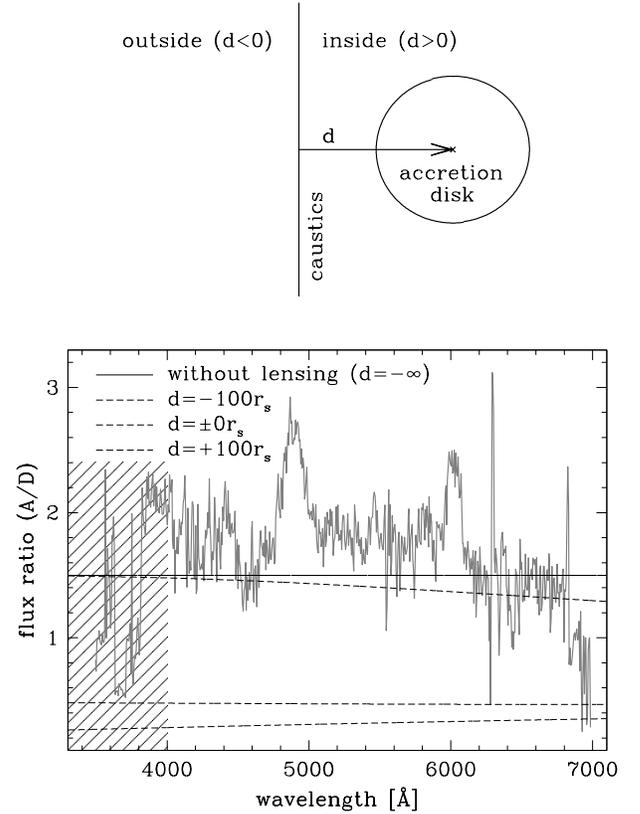}
	\caption{\label{microln} Top panel: Schematic of the straight line caustic with respect to the source. d is the distance measured between the center of the source and the caustic. Bottom figure: flux ratio between \he\ images A and D. Superimposed on the flux ratio observed curve, the straight lines are the modeled flux ratios in different situations, that is for different distances d. The solid line is the modeled flux ratio without microlensing (i.e. d=$\infty$). The dotted, the dashed, and the long-dashed lines correspond, respectively, to the modeled flux ratios with microlensing for distances between the caustic and the source center as described in the legend. The shaded area denotes the blue extremum of the grism (untrustable region).}
\end{figure}

\begin{figure}
	\centering
	\includegraphics[width=8cm,bb=0 0 422 324]{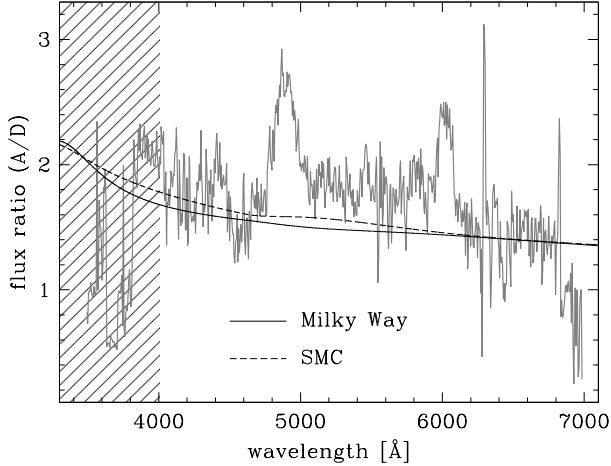}
	\caption{\label{dustex}A normal extinction law, $R_V=3.1$, for dust in the Milky Way (solid line) and an extinction law as in the Small Magellanic Cloud (dotted line). The shaded area denotes the blue extremum of the grism (untrustable region).}
\end{figure}

	Microlensing is not the only possible explanation for the chromatic effect. Galactic dust extinction through the lensing galaxy is also an alternative. In this case, galactic dust would block the blue light from image D, making this image look redder in comparison to image A, and resulting in a decreasing flux ratio towards the redder part of the spectrum, as observed between images A and D. We evaluate the chromatic feature which is to be expected from differential dust extinction \cite[e.g.,][]{jean98}. For these calculations, we consider a normal extinction law, $R_V=3.1$, for dust in the Milky Way \cite[]{cardelli89} and an extinction law as in the Small Magellanic Cloud \cite[]{gordon03}. In both cases, we assume a local gas-to-dust ratio \cite[]{bohlin78} $A_V=5.3\times10^{-22}N_{H}$, where $A_V$ and $N_{H}$ are, respectively, the dust extinction in the $V$ band (in magnitudes) and the column density of hydrogen gas in ${\rm cm^{-2}}$. We also assume that the difference in the gas column density in front of the images is $5\times10^{20} {\rm cm^{-2}}$ in the case of the Milky Way, and is $5\times10^{21}{\rm cm^{-2}}$ in the case of Small Magellanic Cloud. Fig. \ref{dustex} shows the extinction expected from these two models. Both of them seem to trace at least part of the slope that we see in the continuum flux ratio and therefore galactic extinction is a plausible explanation for the phenomenon.

	Thus, in the case of \he, our preferred interpretation for the difference between the spectra of images A and D is microlensing due to the lensing galaxy LG1 and affecting image D, as it explains in the spectrum of image D, both its emission line dimming and its continuum reddening. However, this is not the only interpretation: in particular the effect of the interstellar matter and dust in the lensing galaxy LG1 should not be overlooked.

\subsection{Spectral differences in \rxj}

	The different images of quasar \rxj\ also exhibit spectral flux differences. However due to the small separation of the images, it is impossible to apply the same extraction procedure as for \he. In order to quantify the differences between the spectra, we have performed Gaussian profile fits to the three main emission lines in the LR-blue wavelength range. Using the continuum levels obtained from the Gaussian fits we have derived the equivalent widths (EW) of the emission lines, and considered them as a measure of their strength. This was performed using:

\begin{center}
\begin{equation}
EW = \int^{\lambda_2}_{\lambda_1} \frac{(S(\lambda)-I)}{I} d\lambda
\end{equation}
\end{center}
where $S(\lambda)$ is the flux at wavelength $\lambda$ and I is the continuum level for the chosen wavelength range (in this case the range of the emission lines, delimited by $\lambda_1$ and $\lambda_2$). The parameters are displayed in Table 3.

\begin{table}
\renewcommand{\arraystretch}{1.0}
\centering
\begin{center}
\caption{\label{ratio} Parameters of the Gaussian fits for the emission lines in the system \rxj: amplitude and continuum level are expressed in $10^{-15}$ $erg$ $s^{-1}$, position centers, Gaussian $\sigma$ and equivalent widths in \AA{}. The equivalent widths are measured in the rest frame.}

{\scriptsize
\begin{tabular}{l l l l l l l }
\hline
\hline
LINE&  \multicolumn{2}{c}{{$Ly \alpha$} }    & \multicolumn{2}{c}{{$Si_{IV}$}} &\multicolumn{2}{c}{{$C_{IV}$} }   \\
\hline
      
IMAGE       & A &B & A &B & A &B   \\
\hline
\hline
Amp. &       4.42  &    0.59  &    0.72  &    0.10   &    1.86   &   0.24 \\
$\lambda_{Cent}$ [\AA{}] &       4639  &     4636  &     5292  &     5287   &    5853   &    5843 \\
$\sigma$ [\AA{}] &       75.95  &     73.69  &     33.03  &     31.63   &    55.61   &    59.35 \\
Cont. &       2.85  &    0.47  &     2.94  &    0.50   &    2.70   &   0.46 \\
EW [\AA{}]&    81.83   &    63.69    &   5.43   &    4.50   &    24.93   &   20.13 \\

\hline
\end{tabular}
}
\end{center}
\end{table}

\noindent The differences between images A and B are confirmed by looking at the equivalent width values which are systematically larger in image A. While the continuum ratio is around 5.9, the emission line ratio goes up to 7.7. Emission line residuals are seen in the flux ratio between images A and B, but in contrast to \he, the emission lines appear to be stronger in the image that is closer to the lensing galaxies. As argued in Sect. 6.1, we think that microlensing is the most probable explanation for the emission line residuals, but in this case, given the distances of the images to the lensing galaxy, we are rather dealing with microlensing demagnification in the image complex A.

	To probe this interpretation, we have built a ``macro'' lens model for \rxj, using an elliptical effective lensing potential \cite[]{blandford87} with external shear. 
	
	Using this lens model, we have evaluated the effective shear ($\gamma_{eff}$) and the convergence ($\kappa{_\star}^{eff}$) on all the images of the quasar, in order to characterize the properties of the Fermat travel-time surface \cite[]{blandford86}: minima, maxima or saddle points. The results are shown in Fig. 15. Since images A1 and A3 are located at saddle points, they are more likely to be demagnified by microlensing \cite[]{schechter02}, therefore it is quite plausible that the continuum emitting region of image complex A (A1, A2, A3) is overall demagnified.

	Of course, the effective shear and convergence depend on the convergence provided by a smooth distribution of matter ($\kappa_c$) and a clumpy stellar component ($\kappa_{\star}$) in the following way:
\begin{center}
\begin{eqnarray}
\kappa_{\star}^{eff} &=& \frac{\kappa{_\star}}{1-\kappa_c}\\
\gamma_{eff} &=&  \frac{\gamma}{1-\kappa_c}
\nonumber
\end{eqnarray}
\end{center}

\begin{figure}
	\centering
	\includegraphics[width=8cm,bb=0 0 542 526]{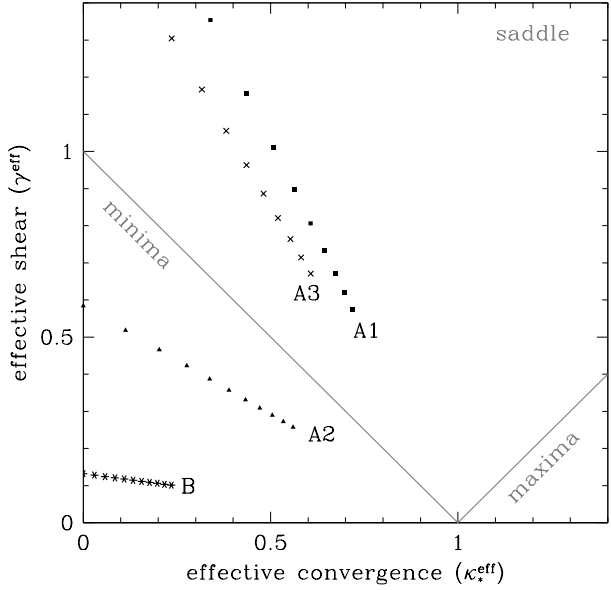}
	\caption{\label{rxjmodel} Results of a simple lens model for \rxj: Different values for the effective shear versus effective convergence. The symbols correspond to different images of the quasar: black squares: A1, black triangles: A2, crosses: A3 and stars: B. From left to right, the contribution of $\kappa_c$ to $\kappa_\star$ varies from 100\% and 0\%. All combinations place the images A1 and A3 in the zone of saddle points.}
\end{figure}

	However, the quasar images A1 and A3 lie at saddle points, no matter which fraction of $\kappa_c$ and  $\kappa_{\star}$ is incorporated in the lens model (see Fig. \ref{rxjmodel}).

\begin{figure} 
	\centering
	 \includegraphics[width=8cm,angle=0]{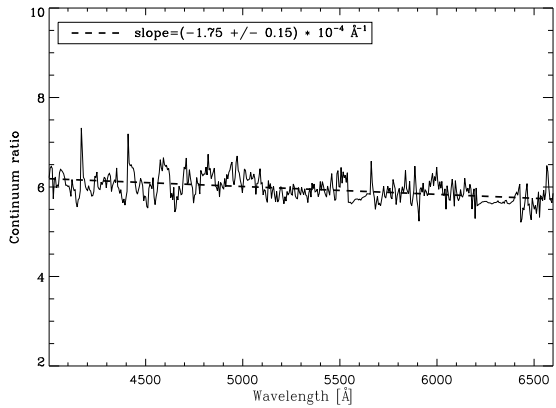}
 	\caption{A linear fit to the continuum ratio for the lensing system \rxj. A mild slope can be seen.}
	\label{conratrx}
\end{figure}

	Regarding the slope of the continuum flux ratio for \rxj, a mild tilt is also observed. We have measured the slope of this variation by adjusting a straight line over the whole spectral range (see Fig. \ref{conratrx}). The fit has been adjusted over the continuum ratio which was obtained after subtracting the Gaussian fits of the emission lines from the total flux ratio. We have measured a slope of  $(-1.75 \pm 0.15) \times10^{-4}$\AA{}$^{-1}$.  Due to the evidence for microlensing demagnification, as argued above to explain the emission line residuals, we are led to consider that microlensing is also responsible for the chromatic distortions, making the complex image A appear bluer. However, this measured effect is of small significance and impossible to disentangle from differential extinction with the present dataset.

\subsection{Spectral differences in \h1413}

\begin{figure}
	\centering
	\includegraphics[width=8cm,bb=0 0 504 361]{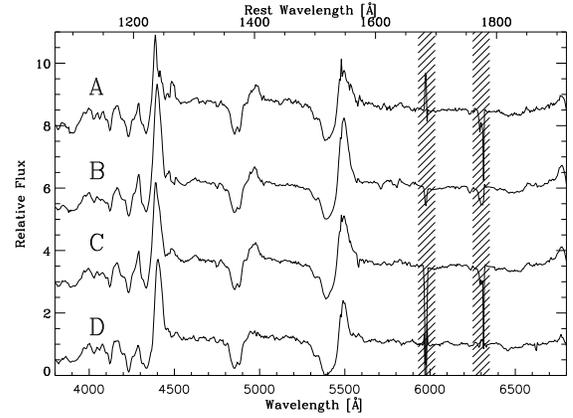}
	\caption{Individual spectra for the four images of the quasar \h1413\ after the fitting procedure. They have been normalized and shifted vertically for easier comparison (by 2.5 with respect to the bottom one). Shaded areas highlight regions affected by sky residuals.}
	\label{spech1413sep}
\end{figure}

	In order to extract an individual spectrum for each image of the quasar we apply the procedure described earlier in Sect 6.1. As before, owing to the small separation of the images, it is possible to do so only with data obtained in the low resolution configuration. The spectra of the four images are displayed in Fig. \ref{spech1413sep}.

\begin{figure}
	\centering
	\includegraphics[width=8cm,bb=0 0 504 360]{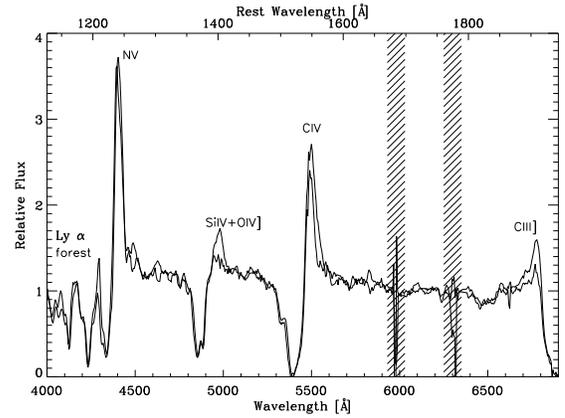}
	\caption{The normalized spectrum of image D of \h1413 (thin line) and the mean spectrum of images A, B and C (thick line). Shaded areas show regions affected by sky line residuals may. The difference in the relative amplification between the continuum and the emission lines is evident, especially at the SiIV and CIV lines.} 
	\label{ratioalld}
\end{figure}

	We notice some differences between the spectra. The quasar image D exhibits the most obvious differences when compared to the other images (see also Fig \ref{ratioalld}): the CIV line clearly shows a smaller intensity with respect to the continuum, while the SiIV+OIV] line has practically vanished. This again is consistent with the continuum of image D being magnified by microlensing.

	\cite{kayser90} suggested microlensing as an explanation for photometric variations seen in image D of the quasar. Using integral field spectroscopy observations, \cite{angonin90} found spectral differences between image D and the other three images. They suggested that the differences were produced by broad absorption line clouds being selectively magnified by microlensing. A comparison between the present VIMOS spectral dataset and the \cite{angonin90} dataset, in particular for image D, shows that the relative spectra have changed: the CIV broad absorption feature 16 years ago was wider in image D than in the other images, whereas it now looks the same, and the SiIV+OIV] emission feature is even smaller in the VIMOS dataset ($\sim$1/3 compared to the average of the other images measured from the continuum) than in the \cite{angonin90} dataset ($\sim$1/2 compared to the average of the other images). In addition, there appears to be a slight slope difference in the \cite{angonin90} dataset when image D is compared to the other images. However, this difference is not observed in the VIMOS dataset. Again some or all of these features could be explained by three phenomena: galactic extinction, microlensing or intrinsic variability. We comment below upon the evidences against or in favor of such possibilities:

\begin{itemize}

	\item \underline{Galactic extinction}: if there is a steeper slope in the continuum of a quasar image when compared to the other images, it can be interpreted as galactic extinction. In the \cite{angonin90} dataset this might be the case with image D, however, no slope difference can be seen seen in the VIMOS dataset. As galactic extinction is not expected to vary over time (especially on such a short time scale of 15 years) the slope should have persisted. Furthermore, \cite{turnshek97} showed that the reddenings between images B and C, as well as between images A and C were considerably greater than that between images D and C. Finally, differential extinction cannot explain the relative flux differences between the continuum and the emission lines when measured in different images. Therefore, we reject this interpretation.

	\item \underline{Microlensing}: the variation of the continuum slope of image D, together with the variation on the relative intensity of the emission lines in all four images (particularly image D) give a clear signal that microlensing is at work and is the best candidate for generating the observed changes. As explained previously, microlensing can explain the chromatic continuum magnification. In this case, the image that suffers magnification of its continuum is also the one that showed a bluer spectrum, which is the most likely kind of chromaticity induced by continuum microlensing. The differences of the broad absorption features seen in \cite{angonin90} were interpreted as microlensing from broad absorption clouds. This suggestion has been tested with different microlensing simulations over the years \cite[e.g.,][]{hutsemekers93,hutsemekers94,lewis98}, simulations in which the BAL profile of image D was reproducible. These differences are not seen any longer in the VIMOS dataset, which again is a result favoring microlensing in image D, as 15 years is roughly the expected timescale for microlensing variations in \h1413 \cite[]{hutsemekers93}. Finally, \cite{ostensen97} have shown, from their photometric monitoring between 1987 and 1995, almost parallel light curves for the quasar images, except for image D. Image D exhibits a slightly higher amplitude in its overall variation, which, added to its spectral differences, makes a strong case for additional microlensing affecting image D.

	\item \underline{Intrinsic variabilit}y: intrinsic variations of the quasar flux coupled with time-delays between its multiple images would also induce spectral differences between the images. Time-delays for this quasar have not been measured yet. Even though its expected value depends on the model, the symmetric disposition of the images suggests that it should be of the order of a month \cite[e.g., model by][]{chae99}. If the timescale of the quasar intrinsic variations is shorter than the time-delay for image D, then the brightness differences could also be explained by this phenomenon and should be seen in all four images at epochs separated by the time-delays.
\end{itemize}

	Although it is not possible today to rule out the quasar intrinsic variability as the cause of the spectral differences between the images (without doing a comprehensive and quantitative study based upon regular and precise spectrophotometric monitoring), we find it very likely that the spectral differences seen in \h1413\ result from microlensing. Microlensing can explain all the feature differences seen in the spectra and is a natural explanation.

\section{Summary and conclusions}\label{conclusion}

We used the VIMOS-IFU device at the VLT to search for objects in the vicinity of four lensed quasars for which a galaxy group or galaxy cluster was either confirmed or expected as part of the lensing system. Additionally, at least two of these lensed quasars were suspected to be under flux (de)magnification due to microlensing. We have described our data reduction techniques for integral field spectroscopy as an aid for future VIMOS users.

We were able to measure flux ratio differences between the different quasar images in three of the systems: \he, \rxj\ and \h1413. Using the high quality spectra collected, our interpretation of the origin of these phenomena is microlensing in the three cases. Based on differences in the flux ratios between the different images when measured in the continuum or the emission lines, we conclude that: \he\ is affected by microlensing magnification on the continuum of image D while image A (A1+A2+A3) of \rxj\ is affected by a microlensing de-magnification of the continuum. Finally image D of \h1413\ shows a consistent history of having been affected by microlensing over the past 15 years. However, even though less likely as described in Sect. 6.3, these flux ratio differences could also be attributed to intrinsic variability from the quasars coupled with time-delays between the images.

In the three cases, the flux ratios between the quasar images show or have shown a non-zero slope, which is also consistent with microlensing. Nevertheless, in the \rxj\ and \he\ systems this phenomenon could also be explained by dust extinction due to the lensing galaxy or, even more likely, a combination of both phenomena. For \h1413, if there has been a slope difference between the images, galactic extinction is not a likely explanation as this slope is not seen in the present VIMOS dataset.

If, as we suspect, the three systems are affected by microlensing, future photometric and/or spectroscopic monitoring should fully characterize the phenomenon and allow derivation of the physical properties of the background sources.

We have also detected absorption features from the two main lensing galaxies in the system \he\ at $z_1$=0.521 and $z_2$=0.523. A third galaxy at similar redshift (z=0.518) has also been detected $\sim$17'' away from the quasar images. In the field around \rxj\ we have newly identified a galaxy, at a distance $\Delta\theta\sim$12'' from the quasar images, and at redshift z=0.484, which makes it independent of the known galaxy group in the line-of-sight to this quasar. Even though not directly bound to the lensing galaxies, its vicinity to the quasar line-of-sight suggests that it might influence the total lensing potential. For \h1413\ the higher resolution spectra clearly reveal the already known absorption systems at redshifts $\sim$1.44 and $\sim$1.66 with high accuracy.

Unfortunately, due to a combination of low exposure time, intrinsic PSF distortions and non-optimal observing conditions, the quality of the data obtained for \b1359\ does not reveal any new information on the system or its surroundings.

\bigskip

\begin{small} 
Acknowledgments. TA and CF are supported by the European Community's Sixth Framework Marie Curie Research Training Network Programme, Contract No. MRTN-CT-2004-505183 ``ANGLES". TA additionally acknowledges support from the International Max
Planck Research School for Astronomy and Cosmic Physics at the
University of Heidelberg. The authors would also like to thank B. Garilli, P. Franzetti and E. Jullo for invaluable advice and help regarding the data reduction and the ESO staff member Burkhardt Wolff for information concerning the VIMOS instrument.

\end{small}

\bibliography{7306}

\end{document}